\documentclass[epjCONF,columns]{svjour}
\usepackage{graphics}
\usepackage[varg]{txfonts} 
\usepackage[latin1]{inputenc}
\session-title{Hadron Collider Physics Symposium 2011}
\begin{document}
\title{New Measurements of Upsilon Spin Alignment at the Tevatron}
\author{Matthew Jones\thanks{\email{mjones@physics.purdue.edu}}
for the CDF collaboration}

\institute{Department of Physics, Purdue University, USA} 
\abstract{
We describe a new analysis of $\Upsilon(nS)\rightarrow \mu^+\mu^-$ decays
collected in $p\overline p$ collisions with the CDF II detector at the
Fermilab Tevatron.  This analysis measures the angular distributions of
the final state muons in the $\Upsilon$ rest frame, providing new information
about $\Upsilon$ production polarization.  We find the angular distributions
to be nearly isotropic up to $\Upsilon$ $p_T$ of $40\;{\rm GeV}/c$,
consistent with previous measurements by CDF, but inconsistent with
results obtained by the D0 experiment.  The results are compared with
recent NLO calculations based on color-singlet matrix elements and
non-relativistic QCD with color-octet matrix elements.}
\maketitle
\section{Introduction}
\label{sec-intro}

A recent analysis~\cite{bib-arxiv} of $\Upsilon\rightarrow\mu^+\mu^-$ decays 
collected with the CDF II detector provides new measurements of the
distributions of decay angles, which depend on the polarization of
$\Upsilon$ states produced in $p\overline p$ collisions.  Previous
measurements~\cite{bib-cdf,bib-d0} carried out by the CDF and D0 experiments
provided useful, but incomplete information about these angular distributions
and did not strongly favor the predictions of any of the
models~\cite{bib-kt,bib-co} used to calculate the production
cross sections.  Furthermore, the fact that these previous measurements
are in apparent disagreement has led to the speculation that
significant acceptance biases could have been overlooked, motivating the
need to perform additional tests of internal consistency in future
measurements~\cite{bib-faccioli}.

The analysis described here measures the full angular distributions of
the final state muons from $\Upsilon(1S)$, $\Upsilon(2S)$, and $\Upsilon(3S)$
decays as functions of
the $\Upsilon$ transverse momentum up to $40\;{\rm GeV}/c$.  This is the
first analysis to report measurements of the $\Upsilon(3S)$ spin alignment.
It is also the first analysis to measure spin alignment in two different
coordinate frames and to compare rotationally invariant quantities in these
frames to demonstrate internal consistency.

\section{Analysis Overview}
\label{sec-overview}
In the rest frame of the $\Upsilon$ decay, the direction of the positive
muon is described using polar angles $(\theta,\varphi)$ measured with respect
to a given set of coordinate axes.  The $s$--channel helicity frame, used in
earlier analyses, defined the $z$-axis along the $\Upsilon$ momentum
vector, with the $x$-axis in the production plane and the $y$-axis
perpendicular to both $x$- and $z$-axes.  Alternatively, the Collins--Soper
frame~\cite{bib-cs} can be used for which the $z$-axis approximates,
on average, the direction of the velocity of the colliding partons.
The form of the angular distribution is constrained by angular momentum
conservation; for a vector meson decaying to two fermions it can be
written
\begin{eqnarray}
  \frac{dN}{d\Omega}&\sim&1 + \lambda_\theta\cos^2\theta + \lambda_\varphi \sin^2\theta\cos 2\varphi + \lambda_{\theta\varphi}\sin 2\theta\cos\varphi + \nonumber \\
 & & \lambda_{\varphi}^\perp \sin^2\theta \sin 2\varphi + \lambda_{\theta\varphi}^\perp \sin 2\theta\sin\varphi.
\label{eqn-dndomega}
\end{eqnarray}
A four-fold symmetry in the acceptance may be exploited to increase
statistics in small bins of solid angle by combining
$(\theta,\varphi)$ with $(\theta,-\varphi)$ and $(\theta,\varphi)$ with
$(\pi-\theta,\pi-\varphi)$, although this leads to a cancellation of the
terms with $\lambda^\perp_\varphi$ and $\lambda^\perp_{\theta\varphi}$
as coefficients.  The coefficients quantify the shape of the angular
distribution and provide direct information about the polarization of the
ensemble of $\Upsilon$ states since they are related to the
elements of the spin density matrix elements $\rho_{ij}$~\cite{bib-noman}
by the expressions $\lambda_\theta=(\rho_{11}-\rho_{00})/(\rho_{11}+\rho_{00})$,
$\lambda_\varphi=\rho_{10}/(\rho_{11}+\rho_{00})$, and
$\lambda_{\theta\varphi}=\rho_{1,-1}/(\rho_{11}+\rho_{00})$.

When a sample of $\Upsilon$ decays is selected using a dimuon trigger, the
observed angular distribution will differ from the form in
Eq.~(\ref{eqn-dndomega}) because of the limited acceptance imposed by
the muon $p_T$ thresholds in the trigger and the geometric coverage of the
detector systems.  The acceptance can change rapidly with both the
transverse momentum and mass of the dimuon system but can be
calculated accurately using a combination of Monte Carlo simulations,
which model the detector geometry, and trigger efficiencies measured using
independent data samples.
The angular distributions of dimuons with mass that include one of the
$\Upsilon(nS)$ resonances will depend strongly on angular distributions
present in
non-resonant backgrounds, which can be highly non-isotropic.
These may have angular distributions that can be parameterized
using Eq.~(\ref{eqn-dndomega}), but they could be more complex
since they do not necessarily arise from the decay of a single vector
state.

\subsection{Previous Analyses}
\label{sec-previous}
Previous $\Upsilon$ spin alignment analyses~\cite{bib-cdf,bib-d0}
were performed using only the $s$--channel helicity frame and
integrated angular distributions over $\varphi$, retaining only sensitivity
to the coefficient $\lambda_\theta$ which is frequently denoted
$\alpha$ in the literature.
This was carried out in several ranges of $p_T(\Upsilon)$ by fitting the
dimuon mass distribution in discrete ranges of $\cos\theta$ to determine
the $\Upsilon$ yields, correcting for detector acceptance, and fitting the
resulting distributions to a function of the form $1+\alpha\cos^2\theta$.
In practice, this is not a trivial procedure because both the $\Upsilon$
acceptance and the shape of the background mass distribution change
significantly with both $p_T$ and $\cos\theta$.

The limitations of these analyses have been pointed out for several
years now~\cite{bib-faccioli}.
The measurement of only one of the three coefficients in
Eq.~(\ref{eqn-dndomega}) does not allow the calculation of rotationally
invariant quantities or the transformation $\lambda_\theta$ from the
$s$--channel helicity frame to different coordinate systems.  This
procedure also does not generalize well to the analysis in many small
bins of $\cos\theta$ and $\varphi$ because the large number of fits to
invariant mass distributions with poorly constrained background shapes
may suffer from large statistical fluctuations or systematic biases.

\subsection{A New Approach}
\label{sec-new}
The analysis procedure described here divides $(\cos\theta,\varphi)$
into $20\times 36$ bins, but it avoids the need to measure
$\Upsilon$ yields in each bin separately.  Instead, all dimuon events
with invariant mass near each of the $\Upsilon$ signals are selected and
the observed numbers of events in each bin are modeled using separate angular
distributions of the form in Eq.~(\ref{eqn-dndomega}) for signal
and background, multiplied by the detector acceptance which is calculated
in each individual bin.  The parameters $\lambda_\theta$, $\lambda_\varphi$,
$\lambda_{\theta\varphi}$ corresponding to the angular distribution of
the $\Upsilon$ signal can then be measured provided the amount of
background and its angular distribution are known.

Within a given range of dimuon $p_T$, the amount of background under
the $\Upsilon(nS)$ signals is determined from a fit to their invariant
mass distribution but an
independent sample is needed to constrain the shape of the angular
distribution from background sources.  Such a sample is obtained by
demanding that the extrapolated trajectory of at least one of the muons
misses the average beam axis by a distance $|d_0|>150\;{\rm \mu m}$.
Although this ``displaced'' muon sample contains a few percent of the
$\Upsilon$ signal due to the $d_0$ measurement resolution, it mostly
selects muons produced in semileptonic decays of heavy quarks, which
forms the dominant source of background.  Since the impact
parameter requirement does not bias the muon decay angle, the angular
distribution of muons from background sources is expected to be the same
in the complementary ``prompt'' muon sample.

A simultaneous fit is then performed to the angular distributions of dimuon
events in prompt and displaced samples selected from ranges of invariant mass
around each of the $\Upsilon$ signals.  In each bin of $p_T(\Upsilon)$,
the fit is performed by maximizing the likelihood function
constructed from the probabilities of obtaining the observed numbers of
events in bins of $(\cos\theta,\varphi)$ given expected yields
calculated using
\begin{eqnarray}
  \frac{dN_p}{d\Omega_{ij}} &\sim&
   N_{\Upsilon} f_p {\cal A}_{\Upsilon}(\cos\theta_i,\varphi_j) \cdot
   w_\Upsilon(\cos\theta_i,\varphi_j;\bar{\lambda}_\Upsilon)+  \nonumber \\
   & & N_d s_p {\cal A}_{b}(\cos\theta_i,\varphi_j) \cdot
   w_{b}(\cos\theta_i,\varphi_j;\bar{\lambda}_{b}), \label{eqn-p} \\
  \frac{dN_d}{d\Omega_{ij}} &\sim&
   N_{\Upsilon} (1-f_p) {\cal A}_{\Upsilon}(\cos\theta_i,\varphi_j) \cdot
   w_\Upsilon(\cos\theta_i,\varphi_j;\bar{\lambda}_\Upsilon) + \nonumber \\
   & & N_d {\cal A}_{b}(\cos\theta_i,\varphi_j) \cdot
   w_b(\cos\theta_i,\varphi_j;\bar{\lambda}_{b}). \label{eqn-d}
\end{eqnarray}
In these expressions, $N_{\Upsilon}$ and $N_{d}$ are the numbers of $\Upsilon$
and displaced background events, $f_p$ is the fraction of the $\Upsilon$
signal retained in the prompt sample, and $s_p$ is the ratio of the
background yields in prompt and displaced samples.  The parameters $f_p$
and $s_p$ are constrained using fits to the prompt and displaced mass
distributions.  The acceptance
${\cal A}_{\Upsilon}$ for $\Upsilon$ signal and ${\cal A}_b$ for dimuon
background events are calculated using Monte Carlo simulations and
the measured trigger and muon selection efficiencies.  The underlying
angular distribution of muons from $\Upsilon$ decays, $w_{\Upsilon}$,
is calculated using Eq.~(\ref{eqn-dndomega}) with coefficients denoted
collectively as $\bar\lambda_{\Upsilon}$.  The angular distribution of
muons in the background component, $w_b$, is similar, but has an additional
term, described below, that allows a better description of the data.

\section{Analysis of CDF Data}
\label{sec-data}
\subsection{Upsilon Trigger}
\label{sec-trigger}
The analysis of $\Upsilon\rightarrow \mu^+\mu^-$ decays is performed at CDF
using a sample of events collected with a 3-level dimuon trigger.  This trigger
required the presence of two oppositely charged tracks at level 1 with
$p_T>1.5\;{\rm GeV}/c$ that extrapolate to hits in one of the CDF muon
detector systems~\cite{bib-xft}.  At least one of the muons had to be in
the central region and the level 2 trigger required that it was also detected
in a second muon detector system located behind additional steel
absorber.  After full event reconstruction, the level 3 trigger required
that this muon had $p_T>4\;{\rm GeV}/c$, the other muon had
$p_T>3\;{\rm GeV}/c$ and that the invariant mass of the pair was between
$8$ and $12\;{\rm GeV}/c^2$.  The geometric acceptance of these triggers
restricts the rapidity of the $\Upsilon$ sample to the central region,
$|y(\Upsilon)|<0.6$.

\begin{figure}
\resizebox{1.0\columnwidth}{!}{
  \includegraphics{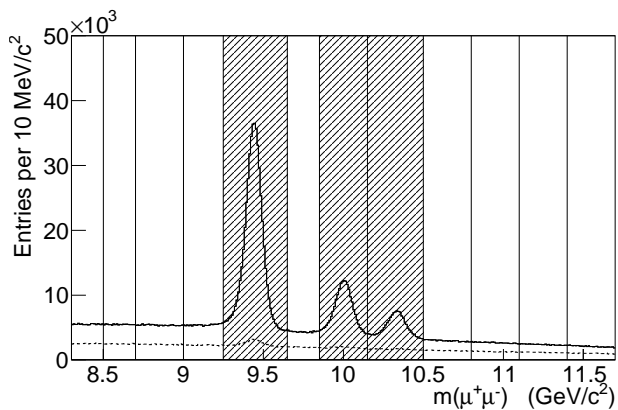} }
\caption{Distribution of $\mu^+\mu^-$ invariant mass.  Pairs where both
originate from close to the beam axis are indicated by the solid histogram,
while those for which at least one misses the beamline by more than
$150\;{\rm \mu m}$ are shown in the dashed histogram.  The shaded regions
indicate the range of masses used to select the $\Upsilon(1S)$, $\Upsilon(2S)$
and $\Upsilon(3S)$ states, while the other regions are used to study
background properties.}
\label{fig-mass}
\end{figure}

For most of Run II, the level 2 trigger was
prescaled dynamically to maintain an approximately constant accept rate and
dead time, but more recently, the level 1 trigger was disabled when
instantaneous luminosities were greater than
$280\times 10^{30}\;{\rm cm}^{-2}{\rm s}^{-1}$.  The prescaled triggers
integrated approximately 70\% of the delivered luminosity, averaged over
the first $6.7\;{\rm fb}^{-1}$ of data collected in Run II.
Figure~\ref{fig-mass} shows the mass distribution of dimuons collected
using these triggers that are used in the angular analysis.

\subsection{Angular Distributions in Background Events}
\label{sec-background}
\begin{figure}
\resizebox{1.0\columnwidth}{!}{
  \includegraphics{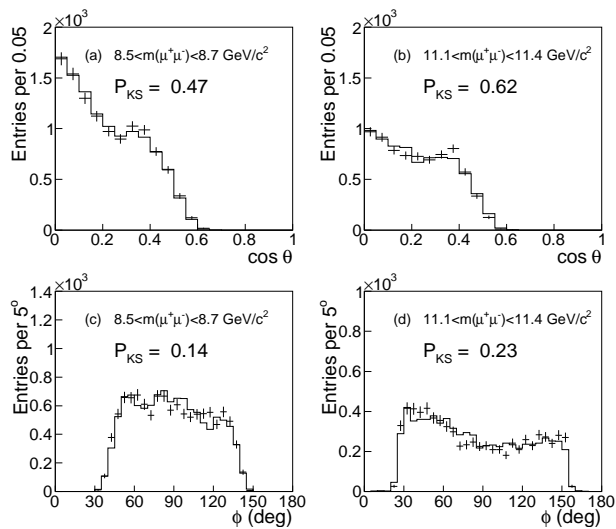} }
\caption{Comparisons of $\cos\theta$ (a,b) and $\varphi$ (c,d) distributions,
measured in the Collins--Soper frame, for prompt and displaced
samples (histograms and error bars, respectively).  Reasonable
agreement, quantified by computing the Kolmogorov--Smirnov statistic,
is observed in mass regions both below (a,c) the $\Upsilon(1S)$ and above
(b,d) the $\Upsilon(3S)$ resonances.}
\label{fig-pdcsphi}
\end{figure}

The analysis of the angular distributions of muons from $\Upsilon$ decays
relies on accurately subtracting the angular distributions that are present
in the background which are estimated using the displaced track sample.
The validity of this procedure is checked by comparing the observed
distribution of decay angles in the high and low mass sideband regions.
The example shown in Fig.~\ref{fig-pdcsphi} demonstrates that this is
the case, with consistency between the two samples tested by computing
the Kolmogorov--Smirnov statistic.

\subsection{Simultaneous Fit in Signal Regions}
\label{sec-signal}
The similarities observed in the angular distributions of prompt and
displaced muon samples at masses both above and below the $\Upsilon(nS)$
resonances support the use of the displaced muon sample to model the
background properties under the $\Upsilon$ signals.  Thus, we apply the
simultaneous fit to prompt and displaced muon samples with mass selected
from each of the three regions containing the $\Upsilon$ signals indicated
in Fig.~(\ref{fig-mass}).  To accommodate the description of the angular
distribution of the background, which is observed to be very non-isotropic,
an additional term proportional to $\cos^4\theta$ is added to
Eq.~(\ref{eqn-dndomega}) to obtain the functiona used for $w_b$.
In addition, a component of the sample that is strongly peaked at large
values of $\cos\theta$ in the $s$--channel helicity frame is removed by
requiring that $|p_T(\mu^+)-p_T(\mu^-)|<(p_T(\mu^+\mu^-)-0.5\;{\rm GeV}/c)$.
This restriction is included in the calculation of the $\Upsilon$ acceptance
but has a negligible effect for $p_T(\Upsilon)>6\;{\rm GeV}/c$.

\begin{figure}
\resizebox{1.0\columnwidth}{!}{
  \includegraphics{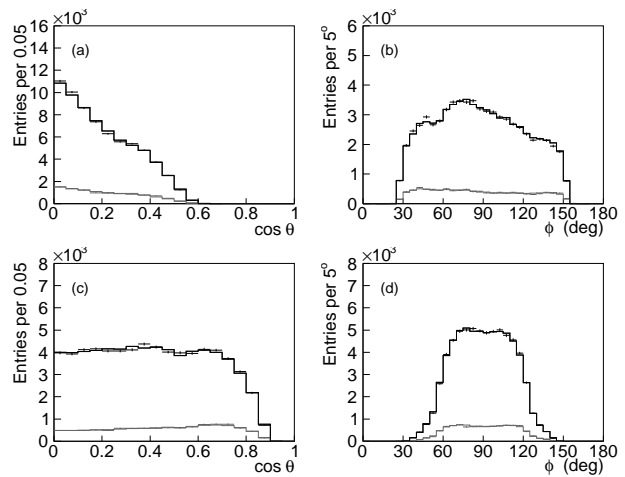} }
\caption{Projections of angular distributions measured in the Collins--Soper
frame (a,b) and the $s$--channel helicity frame (c,d) for muon pairs with
$4<p_T<6\;{\rm GeV}/c$ and mass in the vicinity of the $\Upsilon(1S)$ signal.
The projected distributions in the data are indicated with error bars for
the prompt (black) and displaced (gray) samples.  Solid histograms show the
projections of the fits.}
\label{fig-1sfit}
\end{figure}

The quality of the resulting fit is assessed by
comparing projections of the angular distributions observed in the data
with the corresponding projections of the fit.  Fig.~\ref{fig-1sfit}
shows an example of projections for muon pairs selected from the mass
range containing the $\Upsilon(1S)$ signal.
The quality of fits applied to the other kinematic regions is similar and
shows no systematic trends that depend on either $p_T(\mu^+\mu^-)$ or their
invariant mass.

\subsection{Results}
\label{sec-results}
\begin{figure}
\resizebox{1.0\columnwidth}{!}{
  \includegraphics{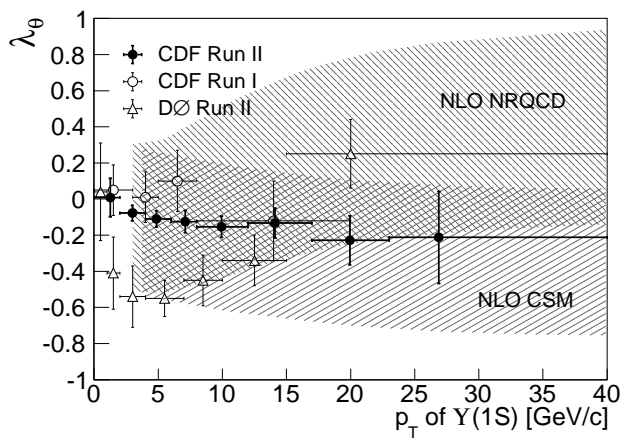} }
\caption{Measurements of $\lambda_\theta$ in the $s$--channel helicity
frame for the $\Upsilon(1S)$ state compared with previous results from
CDF (with $|y(\Upsilon)|<0.4$) and D0 (with $|y|<1.8$) and with
next-to-leading order calculations based on NRQCD with color-octet matrix
 elements, and a next-to-leading order color-singlet model.}
\label{fig-compare}
\end{figure}

The new measurements of $\lambda_\theta$ for 
the $\Upsilon(1S)$ state can be compared with previous results from
the CDF and D0 experiments and with recent NLO
predictions~\cite{bib-qwg,bib-nrqcd,bib-csm}.
These are shown in Fig.~\ref{fig-compare} from which it is apparent that
although the new results are consistent with the Run~I CDF measurement,
they are inconsistent with D0 analysis, with the significance
estimated to be approximately $4.5\sigma$.
The theoretical predictions are currently somewhat imprecise 
due to the poorly measured production cross sections for
$\chi_{bJ}(nP)$ states that decay into $\Upsilon(1S)$~\cite{bib-cdfchib},
but recent results
from the ATLAS experiment\cite{bib-atlas} may improve on this situation.

\section{Rotational Invariants}
\label{sec-rot}
We also demonstrate the internal consistency of the results by
calculating the rotational invariant
$ \tilde\lambda = (\lambda_\theta+3\lambda_\varphi)/(1-\lambda_\varphi)$
using the values of $\lambda_\theta$ and $\lambda_\varphi$ measured in
the $s$--channel helicity frame and the Collins--Soper frames.  Agreement
between the values calculated in each coordinate frame is an important
consistency test because poor determination of the experimental acceptance
or inaccuracies in the subtraction of the highly non-isotropic backgrounds
would be expected to introduce coordinate frame dependent biases in the
measured angular distributions.

The value of $\tilde\lambda$ quantifies the shape of the angular
distribution independent of its orientation with respect to a
coordinate frame.  Decays of $\Upsilon$ states with pure transverse
polarization yield angular distributions with $\tilde\lambda=+1$ while
a purely longitudinal polarization gives $\tilde\lambda=-1$.  A value of
$\tilde\lambda=0$ correspond to an isotropic angular distribution, which
cannot result from the decay of a pure spin-1 state but instead would indicate
that multiple production mechanisms are present leading to an effectively
unpolarized ensemble of decays. 

\begin{figure}
\resizebox{1.0\columnwidth}{!}{
  \includegraphics{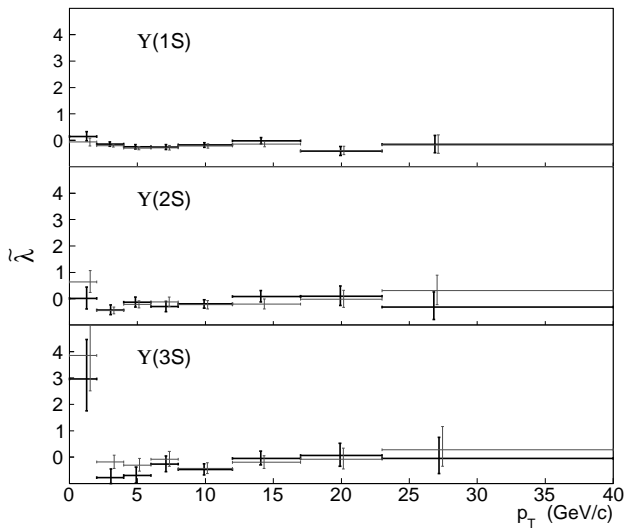} }
\caption{Rotational invariant $\tilde\lambda$, measured in the
Collins--Soper frame (dark lines) and $s$--channel helicity frame (gray lines)
as functions of $p_T(\Upsilon)$.}
\label{fig-lt}
\end{figure}

Figure~\ref{fig-lt} shows the values of $\tilde\lambda$ measured for the
$\Upsilon(1S)$, $\Upsilon(2S)$ and $\Upsilon(3S)$ states as functions of
$p_T(\Upsilon)$ in both the Collins--Soper and the $s$--channel helicity
frames.
Since $\tilde\lambda$ is measured in each coordinate frame using the same
data samples, the statistical uncertainty of each measurement is
highly correlated.  The sizes of variations in values of $\tilde\lambda$
measured in the two frames that would be expected from purely statistical
fluctuations were estimated using Monte Carlo simulations and found to be
generally consistent with the observed differences.  This is the first
analysis to perform such a test and based on these findings, there appears
to be no evidence for significant biases in the calculated acceptance.

The values $\tilde\lambda\approx 0$ reached at large $p_T$ suggest that all
three of the $\Upsilon(nS)$ states are produced in an unpolarized mixture.
This is the first measurement of angular distributions in $\Upsilon(3S)$
decays and is significant because it had been thought that a greater fraction
of $3S$ states should be produced directly, rather than via feed-down from
$\chi_b$ states, in which case the calculated spin alignment predictions
should be more precise.

\section{Conclusions}
\label{sec-conclusions}
The measurements described here provide the most detailed characterization
of the angular distributions of $\Upsilon\rightarrow\mu^+\mu^-$ decays
produced at a hadron collider to date.  We find little evidence for strong
polarization of any of the three $\Upsilon(nS)$ states in the central
region of rapidity $|y|<0.6$ and with $p_T$ up to $40\;{\rm GeV}/c$.
This is consistent with the results previously obtained in Run I by CDF
and inconsistent with measurements carried out by D0 in Run II.  Although
the D0 measurements were carried out over the wider range of rapidity
$|y(\Upsilon)|<1.8$, we find no evidence that the angular distributions
change rapidly in the central region of rapidity accessible to the
CDF detector.  We look forward to the possibility of refined predictions
from theory and new results from the LHC experiments which may be able
to further clarify the experimental situation.

\end{document}